\documentclass[prb,twocolumn,showpacs]{revtex4}
\usepackage{amsmath}
\usepackage{graphicx}
\usepackage{bm}

\newcommand{\be}{\begin{equation}}
\newcommand{\ee}{\end{equation}}

\newcommand{\tb}{Tb$_2$Ti$_2$O$_7$}
\newcommand{\er}{Er$_2$Ti$_2$O$_7$}
\newcommand{\yb}{Yb$_2$Ti$_2$O$_7$}
\newcommand{\ho}{Ho$_2$Ti$_2$O$_7$}
\newcommand{\re}{R$_2$Ti$_2$O$_7$}
\newcommand{\dy}{Dy$_2$Ti$_2$O$_7$}

\begin{document}

\title{Effective spin-1/2 exchange interactions in \tb}

\author{S. P. Mukherjee\footnote{Current address: Department of Physics,
Brock University, 500~Glenridge Avenue, St. Catherine's, Ontario, Canada} and S. H. Curnoe}
\email[Electronic address: ]{curnoe@mun.ca}
\affiliation{Department of Physics and Physical Oceanography,
Memorial University of Newfoundland, St.\ John's, Newfoundland \& Labrador 
A1B 3X7, Canada}

\begin{abstract}
We derive an effective spin-1/2 exchange model 
for non-Kramers Tb$^{3+}$ states
in the pyrochlore \tb.  The four anisotropic nearest-neighbour exchange constants, as
well as  
next-neighbour exchange constants are derived for the effective model.
This work goes beyond the independent tetrahedra model by considering all
nearest-neighbour exchange paths on the pyrochlore lattice. Estimates of
the exchange constants reveal that \tb\ 
is described by a quantum spin ice Hamiltonian.
\end{abstract}


\maketitle

\section{Introduction}

The rare-earth pyrochlore magnets, with chemical formula \re,
exhibit a variety of low-temperature phenomena, from magnetic ordering in 
\er\ [\onlinecite{champion2003}],  to spin ice states in \ho\ [\onlinecite{harris1997}] and \dy\
[\onlinecite{ramirez1999}], and 
possible
spin liquid behaviour in \tb\ [\onlinecite{gardner2001}]. 
In each case, the magnetic properties are due to the magnetic moments of
the rare earth ions, 
which are proportional to 
$J$, the total angular momentum of the ion,
and to interactions between them.
In spite of rather large values of $J$ derived from Hund's rules,
the rare-earth pyrochlores are essentially quantum magnets: 
a strong crystal electric field (CEF) lowers the $2J+1$-fold degeneracy 
of the rare earth ions into
singlets and doublets, with rather large energy differences between the 
levels.   In many cases, the CEF ground state is a doublet, 
which is treated as a basic two-level quantum mechanical system.

Recently, there has been a great deal of effort to 
describe these various rare earth pyrochlores
within the same phenomenological model, the spin-1/2 nearest-neighbour exchange
interaction.  This modeling process is straightforward for materials such as
\er\ and \yb, whose CEF ground state doublets are in fact spinors.  
However, \tb\ has proven to be especially difficult to model, for two 
reasons.  First, the CEF ground state doublet of \tb\ is not a spinor.  
Second,
\tb\ is complicated by the presence of a low-lying CEF excited state just
17.9 K above the ground state, which tends to mix into the ground state
because of the exchange interaction.

The rare earth ions in pyrochlore crystals
 are located at the 16d Wyckoff position of the space
group $Fd\bar{3}m$.   There are four 16d sites in the 
{\em primitive} unit cell, located on the vertices of a tetrahedron.
The local
site symmetry (CEF symmetry) is $D_{3d}$.  
The 3-fold ($C_3$) axes point 
in different directions for the different sites:
$[111]$, $[1\bar 1\bar 1]$, $[\bar 1 1 \bar 1]$ and $[1 \bar 1 \bar 1]$ for sites \#1, 2, 3 and 4
respectively.
These directions define a local $z$-axis for each site
on a tetrahedron.
For non-Kramers ions (such as Tb$^{3+}$ or Ho$^{3+}$),
the CEF states are singlets or doublets 
belonging to the $A_1$, $A_2$ or $E$
representations of $D_3$. For Kramers ions (such as Er$^{3+}$, Yb$^{3+}$ or
Dy$^{3+}$), the CEF states are doublets belonging to either 
the $\Gamma_4$ or $\Gamma_5$ representations of $D_3'$, the double group of 
$D_3$.  
The CEF ground states for Er$^{3+}$ in \er\ and Yb$^{3+}$ in \yb\
belong to 
$\Gamma_4$, which is  isomorphic to spin-1/2.    Therefore 
CEF ground state doublets  of Er$^{3+}$ and Yb$^{3+}$ 
can be easily mapped to a spin-1/2 spinor by 
an appropriate renormalisation of the matrix elements for
the operators $J_z$
and $J_{\pm}$.   
Here we are concerned with finding a map
between the non-Kramers ($E$) doublet  and a spin-1/2 ($\Gamma_4$) doublet.
Because these two kinds of doublets 
transform differently under rotations, such a map must be
constructed with care.  In fact, a symmetry-preserving map
exists
if these doublets are considered in groups of 4 (the four 
vertices of a tetrahedron in the pyrochlore lattice).

In the following section, we describe the CEF ground state
of \tb, and the map between \tb\ and spin-1/2 single
tetrahedron states is defined.
The exchange interaction is treated in Section III, for the general
case, the spin-1/2 case, and for \tb.
A map between spin-1/2 and \tb\ exchange models is given in 
Section IV.  The magnetisation is
discussed in Section V. 
Section VI contains concluding remarks.

\section{Quantum mechanical states for \tb}
\subsection{CEF ground state for Tb$^{3+}$ in \tb}
The CEF Hamiltonian for the rare earth sites in \re\ is given by 
\begin{equation}
H_{\rm {CEF}} = B^0_2O^0_2 + B^0_4 O^0_4 + B^3_4 O^3_4 + B^0_6 O^0_6 
+B^3_6 O^3_6 + B^6_6 O^6_6
\label{HCEF}
\end{equation}
where $O^i_j$ are Stevens operators, $j$-th order  
polynomials of the operator
$\vec{J}$, the total angular momentum. $B^i_j$ are constants
determined experimentally.   There have been several determinations of 
$B^i_j$ for \tb; all\cite{gingras2000,mirebeau2007,bertin2012} but the most recent\cite{zhang2014} are consistent with each other.
The differences in these constants do not
affect the symmetries of the CEF states, but they are
eventually reflected in the 
exchange constants (see Section V).

According to Hund's rules, the total angular momentum of the
Tb$^{3+}$ ion is $J=6$.  The CEF lifts the 13-fold 
degeneracy into singlets and doublets.
The CEF ground state of Tb$^{3+}$ in \tb\ is a doublet,\cite{bertin2012}
\begin{equation}
|\pm\rangle = \pm 0.263|\pm 5\rangle - 0.131 |\pm 2\rangle \mp 0.128|\mp1\rangle-0.947|\mp4\rangle
\label{CEFground}
\end{equation}
where the quantisation axis points along the $C_3$
axis, which points in a different direction
at each site.  The quantisation axis defines a local $z$-axis. 
%Local $x$ and $y$ axes are chosen to be perpendicular to $z$ and to obey the right-hand rule.
In this way, a different set of local axes is defined for each site on
a tetrahedron (see Appendix A for a detailed description).
The matrix elements for the operators $J_{\pm} \equiv J_x \pm i J_y$
within the $|\pm\rangle$ doublet 
are zero, while 
\begin{equation}
j_1 \equiv \langle +|J_z|+\rangle
=-3.21.   \label{j1}
\end{equation}
Here $x,y,z$ subscripts are used to denote local axes, while 
superscripts will be used to denote global axes.

\subsection{The first excited CEF state for Tb$^{3+}$ in \tb}

In \tb, the first excited CEF state (also a doublet) lies only $\Delta = 17.90$~K 
above the ground state.  Therefore, as was recognised long ago,\cite{aleksandrov1981,aleksandrov1985} 
there is a  significant admixture of this excited state to the 
lowest energy states.   However, the symmetry of the lowest energy states
cannot be affected by this admixture.

The first excited CEF state is\cite{bertin2012}
$$
|\uparrow/\downarrow\rangle = \mp 0.923 |\pm 5\rangle 
+ 0.251 |\pm 2\rangle \mp 0.082 |\mp1\rangle 
-0.280 |\mp 4\rangle
$$
The matrix element for $J_z$ within this doublet is 
\begin{equation}
j_2 \equiv \langle \uparrow |J_z|\uparrow\rangle  = 4.05,
\end{equation}
 and the matrix
elements for $J_{\pm}$ are again zero.
The mixing of the first CEF excited state to the ground state 
will depend on the  matrix elements
\begin{equation}
j_3 \equiv \langle \uparrow |J_z|+\rangle = -2.37
\end{equation}
and
\begin{equation}
t \equiv \langle \uparrow |J_{+}| -\rangle = 4.72.
\label{t}
\end{equation}

\subsection{Map between states}
With four sites per tetrahedron, there are 
sixteen \tb\ tetrahedron states of the form  
$|\pm\pm\pm\pm\rangle_{\rm Tb}  \equiv |\pm\rangle_1
\otimes |\pm\rangle_2 \otimes |\pm\rangle_3 \otimes |\pm \rangle_4$,
where $|\pm\rangle_i$ is a non-Kramers doublet on the  $i$th site.  
In a similar fashion, we can also define 
sixteen spin-1/2 tetrahedron states, which will  be denoted as
$|\pm\pm\pm\pm\rangle_{\frac{1}{2}}$.
Each of these kets represents a classical state where each spin can be visualised as 
pointing into or out of the tetrahedron.
There are two anti-ferromagnetic states,
$|----\rangle$ and $|++++\rangle$,
with all four spins pointing into ($-$)  or out of $(+$)  the tetrahedron (``all-in/all-out" states),
while the six states of the form $|++--\rangle$ are ferromagnetic, with two 
spins pointing in and two spins pointing out of the tetrahedron (``2-in-2-out" 
spin ice 
states).   In addition, there are eight ``3-in-1-out/1-in-3-out" states.
 
The symmetry group of a tetrahedron in the pyrochlore lattice
is $T_d$, with representations $A_1$, $A_2$, $E$, $T_1$ and $T_2$.  Both the 
\tb\  and the spin-1/2 tetrahedron states
can be used as basis functions to generate a (reducible) representation of $T_d$.  In both cases, {\em in spite of different transformation properties of 
the individual site states},  the decomposition is $A_{1} \oplus 3 E \oplus 2 T_1 \oplus  T_2$.
This finding allows us to define a map between  the 
\tb\ (non-Kramers) tetrahedron
states and spin-1/2 tetrahedron states.
The map between the \tb\ and spin-1/2
tetrahedron basis states is 
\begin{equation}
|\pm\pm\pm\pm\rangle_{{\rm Tb}} \sim  (-1)^{\eta} {\cal K} \left|\pm \pm 
\pm \pm \right\rangle_{\frac{1}{2}} 
\label{eq:map}
\end{equation}
where ${\cal K}$ stands for time reversal (represented in the 
standard way as $-i\sigma_y {\cal K}_0$, where ${\cal K}_0$ is
complex conjugation)
and the exponent $\eta=0$ for the 2-in-2-out states and the
3-in-1-out spin-1/2 states (but not the 1-in-3-out states);
$\eta=1$ otherwise.  The phase $(-1)^{\eta}$ 
is a reflection of the non-triviality
of the map.

It is worth noting that the tetrahedron states
formed from the third kind of doublet (belonging
to the $\Gamma_5$ representation of $D_3'$) generate a representation with decomposition
$3A_1 \oplus 2 A_2 \oplus E \oplus         
2T_1 \oplus T_2$. Therefore there is no map between 
$\Gamma_5$ tetrahedron states and 
spin-1/2 tetrahedron states that is generally valid.
The CEF ground state of Dy$^{3+}$ in \dy\ belongs to this case.

\section{The exchange interaction}
\subsection{Nearest neighbour exchange interaction: general}
The exchange interaction is a phenomenological model
that describes the 
energy dependence of different relative orientations of
neighbouring magnetic moments.
The general form of the exchange Hamiltonian is governed by
the symmetry of the crystal.  In a highly symmetric crystal, the number
of free parameters of the Hamiltonian is small. 
%These parameters are commonly determined by experiments, and in principle may be calculated using numerical methods and detailed microscopic models.  

The most general form of the of the nearest neighbour exchange interaction on 
the pyrochlore lattice is\cite{curnoe2008}
\begin{equation}
H_{\rm ex} = {\cal J}_1 X_1 + {\cal J}_2 X_2 + {\cal J}_3 X_3 
+{\cal J}_4 X_4
\label{Hex}
\end{equation}
where ${\cal J}_i$ are four independent exchange constants.
It is convenient to express the exchange terms $X_i$ using the 
local axes introduced in the previous section and described in detail in
Appendix A,  
\begin{eqnarray}
X_1 & = & -\frac{1}{3} \sum_{\langle i j\rangle} J_{iz}J_{jz} \label{X1} \\
X_2 & = & -\frac{\sqrt 2}{3} \sum_{\langle i j\rangle} 
[\Lambda_{s_is_j}(J_{iz}J_{j+} + J_{jz}J_{i+}) +\rm{H.c.}]\\
X_3 & = & \frac{1}{3} \sum_{\langle i j\rangle} (\Lambda_{s_is_j}^{*} J_{i+}J_{j+}
+{\rm H.c.}) \\
X_4 & = & -\frac{1}{6} \sum_{\langle i j\rangle} (J_{i+} J_{j-} + \rm{H.c.})
\label{X4}
\end{eqnarray}
where H.c. stands for ``Hermitian conjugate," $\Lambda_{12} = \Lambda_{34}=1$ 
and 
$\Lambda_{13}=\Lambda_{24} = \Lambda_{14}^{*}= \Lambda_{23}^{*}
=\varepsilon \equiv \exp\left(\frac{2\pi i}{3}\right)$.
The sums are over pairs of nearest neighbours and are infinite; the phases
$\Lambda_{s_is_j}$ depend on the site numbers of the 
neighbouring spins.
When $i$ and $j$ are nearest neighbours, the site numbers
$s_i$ and $s_j$ are always different.
Note that in the special case
when
${\cal J}_1={\cal J}_2 = {\cal J}_3 = {\cal J}_4 \equiv {\cal J}$, the exchange 
interaction is isotropic, $H_{\rm ex} = H_{\rm iso} = {\cal J} 
\sum_{\langle i,j\rangle} \vec J_i \cdot \vec J_j$. 

The simplest case is when ${\cal J}_{2,3,4}=0$.   Then the eigenstates of $H_{ex}$ are 
classical states in which every spin is parallel to its
local $z$-axis, pointing 
either into or out of each tetrahedron.   Since each spin sits on the vertex of two vertex-sharing 
tetrahedra, a spin which points out of one tetrahedron necessarily points into the other.
When ${\cal J}_1 >0$,  the ground state is doubly
degenerate: all four spins 
point into or out of each tetrahedron in the lattice.  When ${\cal J}_1 <0$,
the ground state is the highly degenerate 2-in-2-out ``spin ice" state.

The coupling constants ${\cal J}_i$ will be reserved for effective spin-1/2 models.  The coupling constants for \tb\ will be denoted ${\cal I}_i$,
\begin{equation}
H^{\rm Tb}_{\rm ex} =  {\cal I}_1 X_1 + {\cal I}_2 X_2 + {\cal I}_3 X_3 +
{\cal I}_4 X_4.
\label{HTbex}
\end{equation}
$H_{\rm ex}$ (\ref{Hex}) and $H_{\rm ex}^{\rm Tb}$ (\ref{HTbex}) have exactly the same form, but
 with different coupling constants. Also,
$H_{\rm ex}$ acts on 
spin-1/2 states, while $H_{\rm ex}^{\rm Tb}$  acts on 
$J=6$ states.
Our goal is to replace $H_{\rm ex}^{\rm Tb}$ by effective spin-1/2 model,
which we will call $H_{\rm eff}^{\rm Tb}$. 
The exchange constants ${\cal J}_i$ in $H_{\rm eff}^{\rm Tb}$ 
will be expressed in terms of the constants ${\cal I}_i$ in $H_{\rm ex}^{\rm Tb}$.

\begin{figure}[h!]
\includegraphics[height=2.4in]{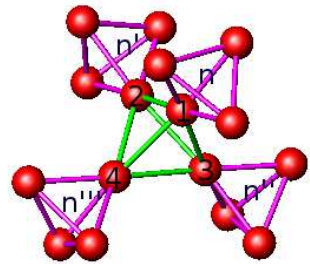}
\caption{(Colour online) The face centred cubic (fcc) unit cell of Tb$_2$Ti$_2$O$_7$, showing only
the Tb$^{3+}$  ions and the exchange paths connecting them.
There are two
orientations of tetrahedra, which we call $A$ (green) and $B$ (pink). 
The lines connecting the
ions (the edges of the tetrahedra) are the exchange paths. 
The labels on the ions (1-4) and the $A$ tetrahedra ($n$, $n'$, $n''$, $n''''$)
illustrate how each ion on a 
$B$ tetrahedron also belongs to an $A$ tetrahedron.
\label{fig1}
}
\end{figure}

We will now analyse the models $H_{\rm ex}$ and $H_{\rm ex}^{\rm Tb}$ in more
detail.
Pairs of nearest neighbours can be visualised as lines that
connect nearest neighbour sites on the lattice.  In the pyrochlore lattice,
these lines are precisely 
the edges of the tetrahedra.  The tetrahedra 
occur in two orientations, $A$ and $B$ (see Fig.\ \ref{fig1}).   
Thus the sum over nearest neighbours can be
split into two parts: the set of all $A$ tetrahedra and the set of all $B$ 
tetrahedra.   
Then the exchange Hamiltonian can be written as  
\begin{equation}
H_{\rm ex} = H^A + H^B
\label{Hex2}
\end{equation}
Let $n$ index the $A$ tetrahedra. For example, $J_{niz}$ is the $J_z$ 
operator (using the local $z$-axis) for the $i$th ($i$=1,2,3,4) site
on the $n$th $A$ tetrahedra.  
Using this notation, we have
\begin{equation}
H^A = \sum_n H^A_n = \sum_n {\cal J}_1 X_{1n}^A + {\cal J}_2 X_{2n}^A
+{\cal J}_3 X_{3n}^A + {\cal J}_4 X_{4n}^A 
\label{HA}
\end{equation}
where, for example, $X_{1n}^A$ is the first exchange term for nearest
neighbours on the $n$th tetrahedron,
\begin{eqnarray}
X_{1n}^A & = &  -\frac{1}{3}(J_{n1z}J_{n2z}+J_{n1z}J_{n3z}+J_{n1z}J_{n4z} \\
& & 
+J_{n2z}J_{n3z}+J_{n2z}J_{n4z}+J_{n3z}J_{n4z}).
\end{eqnarray}

The operator $H^B$ can also be expressed as a sum over A tetrahedra.  
Consider a particular B tetrahedron.  Each of its four spins are located
on the vertices of different 
A tetrahedra.  If the spin on site \#1
is on the $n$-th A tetrahedron, the spin on site \#2 is on the $n'$-th
A tetrahedron, where $n' \equiv n-(1/2,1/2,0)$, the spin on site \#3 is on 
$n''\equiv n-(1/2,0,1/2)$, and the spin on site \#4 is on $n'''\equiv n-(0,1/2,1/2)$, as shown in  Fig.\ \ref{fig1}.
Then
\begin{equation}
H^B = \sum_n H^B_n = \sum_n {\cal J}_1 X_{1n}^B + {\cal J}_2 X_{2n}^B
+{\cal J}_3 X_{3n}^B + {\cal J}_4 X_{4n}^B 
\label{HB}
\end{equation}
where, for example,
\begin{eqnarray}
X_{1n}^B &=& -\frac{1}{3}(J_{n1z}J_{n'2z}+J_{n1z}J_{n''3z}+J_{n1z}J_{n'''4z} 
\nonumber \\
&+ & 
J_{n'2z}J_{n''3z}+J_{n'2z}J_{n'''4z}+J_{n''3z}J_{n'''4z}).
\end{eqnarray}

The exchange Hamilton for \tb, $H^{\rm Tb}_{\rm ex}$ (\ref{HTbex}), can be 
split into A and B parts in a similar way,
\begin{equation}
H_{\rm ex}^{\rm Tb}  =  H^{\rm A,Tb} + H^{\rm B,Tb}.
\label{HexTb2}
\end{equation}

\subsection{The spin-1/2 independent tetrahedra model}
First we consider a simple model involving the exchange paths on a single
tetrahedron (the $n$th A tetrahedron),
\begin{equation}
H^A_n = {\cal J}_1 X_{1n}^A + {\cal J}_2 X_{2n}^A
+{\cal J}_3 X_{3n}^A + {\cal J}_4 X_{4n}^A.
\label{HAn}
\end{equation}
This can be represented as a $16\times 16$ matrix, which can be block diagonalised using
the kets described in Appendix~B.  Exact eigenfunctions can
easily be found. 

The solutions to $H^A$ (\ref{HA}) are the direct product (over
tetrahedra) of the single tetrahedron solutions of $H^A_n$~(\ref{HAn}), which is 
why it is called the ``independent tetrahedra model". 
Because it is exactly solvable, $H^A$ is often used to model
experiments instead of the full Hamiltonian $H_{\rm ex}$
(\ref{Hex}).\cite{molavian2007,dalmas2012,curnoe2013}
$H^A$ is a model that omits half of the exchange paths in
$H_{\rm ex}$, which suggests that the exchange constants of $H^A$ 
are approximately twice as large as those of $H_{\rm ex}$.\cite{dalmas2012} 
We also note that 
$H^A$ has a lower symmetry than $H_{\rm ex}$: instead of the full space
group $Fd\bar{3}m$, it is $F\bar{4}3m$, with point group $T_d$ instead 
of $O_h$.\cite{curnoe2008}

\subsection{The exchange interaction for \tb}
\subsubsection{The exchange interaction for 
non-Kramers spins restricted to the CEF ground state}
When non-Kramers spins are restricted to the CEF ground state, 
the exchange interaction is greatly simplified because the matrix elements
for $J_{\pm}$ vanish within this restriction.
Then the eigenvectors 
are the classical states $|\pm\pm\pm\pm\pm\pm\ldots\rangle$.
The ground state is either the doubly degenerate all-in-all-out state
(${\cal I}_1 > 0$) or a highly degenerate spin ice state (${\cal I}_1 <0$).
% with eigenvalues $-2 j_1^2 {\cal I}_1$ for the states $|++++\rangle$ and $|----\rangle$, $2 j_1^2{\cal I}_1/3$ for the six 2-in-2-out states and $0$ for the eight remaining states.
This model maps to a spin-1/2 model with ${\cal J}_1
= 4 {\cal I}_1 j_1^2$ and ${\cal J}_{2,3,4}=0$.  
This model describes the spin ice material
\ho, but it is insufficient for \tb, for which higher CEF levels
must be included.

\subsubsection{The exchange interaction for non-Kramers spins restricted
to the CEF ground state and first excited state}
Perturbation theory is used to determine the mixing of 
the first excited CEF level to the CEF ground state manifold.  
The unperturbed Hamiltonian is $H_{\rm {CEF}}$  (\ref{HCEF})
restricted to the CEF ground and first excited states, while  the
exchange interaction $H^{\rm Tb}_{\rm ex}$ (\ref{HTbex}) is the perturbation.

Second order perturbation theory yields
an effective exchange Hamiltonian restricted to the CEF ground state:\cite{messiah}
\begin{equation}
H^{\rm Tb}_{\rm eff} = PH^{\rm Tb}_{\rm ex} P + P H^{\rm Tb}_{\rm ex} \frac{Q}{a} H^{\rm Tb}_{\rm ex} P
\label{HTbeff}
\end{equation}
where $P$ is the projector to the CEF ground state and $Q$ 
is the projector that is supplementary to $P$ {\em i.e.} it 
projects states that have one or more spins in the CEF first excited state.  The denominator
$a$ is the energy difference between the ground and excited states.
$P$ is the direct product of projectors $P_n$ which operate on
single tetrahedra.  $Q$ can  also be expressed in terms of 
single tetrahedron operators: on the $n$th tetrahedron, one, two,
three or four spins can be excited, corresponding to the 
projections $Q_{n,{\rm one}}$, $Q_{n,{\rm two}}$ {\em etc.}  
However, for second order perturbation theory, 
we need only consider contributions to $Q$ where
one or two spins are excited because $H^{\rm Tb}_{\rm ex}$ is bilinear in
the spin operators and can only excite up to two spins at a time
via the operators $J_{\pm}$ and $J_z$.
Therefore,
\begin{equation}
\frac{Q}{a} = \sum_n\frac{Q_{n,{\rm one}}}{-\Delta} + \sum_n\frac{Q_{n,{\rm two}}}{
-2 \Delta} + \sum_{n,m<n}\frac{Q_{n,{\rm one}} Q_{m,{\rm one}}}{-2 \Delta}
\end{equation}
where the first term has one spin excited on one tetrahedron,
the second term has two spins excited on one tetrahedron
and the third term has two spins excited on two different tetrahedra.
The operators $Q_{n,{\rm one}}$ and $Q_{n,{\rm two}}$ can be
further expanded as
\begin{eqnarray}
Q_{n,{\rm one}} & = & Q_{n1}P_{n2}P_{n3}P_{n4} + P_{n1}Q_{n2}P_{n3}P_{n4} +  \nonumber \\
& & P_{n1}P_{n2}Q_{n3}P_{n4} + P_{n1}P_{n2}P_{n3}Q_{n4}\\
Q_{n,{\rm two}} & = & Q_{n1}Q_{n2}P_{n3}P_{n4} + Q_{n1}P_{n2}Q_{n3}P_{n4}+ \nonumber \\
&&Q_{n1}P_{n2}P_{n3}Q_{n4} + P_{n1}Q_{n2}Q_{n3}P_{n4}+ \nonumber \\
&&P_{n1}Q_{n2}P_{n3}Q_{n4}+ P_{n1}P_{n2}Q_{n3}Q_{n4}. 
\end{eqnarray}

In Section IV, we show that $H^{\rm Tb}_{\rm eff}$ (\ref{HTbeff}) has the same matrix 
representation as $H_{\rm ex}$ (\ref{Hex}).
However, before considering the full lattice
exchange $H_{\rm eff}^{\rm Tb}$, we will study the
simpler independent tetrahedra model.

\subsubsection{The exchange interaction for non-Kramers spins
in the independent tetrahedra model}
In the independent tetrahedra model, the Hamiltonian 
for Tb$^{3+}$ spins is $H^{A,{\rm Tb}}$.
Perturbation theory yields 
\begin{eqnarray}
H^{A,{\rm Tb}}_{\rm eff} & = &  P H^{A,{\rm Tb}} P + P H^{A,{\rm Tb}} \frac{Q}{a} H^{A,{\rm Tb}} P  \\
& = & \sum_n P H^{A,{\rm Tb}}_{n} P + \sum_{n,m} P H^{A,{\rm Tb}}_{ n}  \frac{Q}{a} H^{A,{\rm Tb}}_m P\\
& = & \sum_n\left[PH^{A,{\rm Tb}}_{n}P \right. \nonumber \\
& & \left. + PH^{A,{\rm Tb}}_n\left(\frac{Q_{n,{\rm one}}}{-\Delta} 
 + \frac{Q_{n,{\rm two}}}{-2\Delta}\right) H^{A,{\rm Tb}}_n P\right] \nonumber \\
& \equiv & \sum_n H_{n, {\rm eff}}^{A,{\rm Tb}}
\label{HATbeff}
\end{eqnarray}
In the second last  line we make use of the fact that $H^{A,{\rm Tb}}_n$ acts only 
within the $n$th tetrahedron, so the only non-zero contribution in
the sum over $n$ and $m$ is when $m=n$.
Therefore this calculation reduces to a single tetrahedron
Hamiltonian $H^{A,{\rm Tb}}_{n,{\rm eff}}$.

\section{Map between spin-1/2 and Tb$^{3+}$ exchange models} 

The results of the single tetrahedron calculation were found previously.\cite{curnoe2013}
By comparing a matrix representation of $H^{A,{\rm Tb}}_{n,{\rm eff}}$ to 
a matrix representation of the spin-1/2 single tetrahedron $H^A_n$, a map between
the Tb$^{3+}$ exchange constants and the spin-1/2 exchange constants was found.
The basis functions that were used to find the matrix 
representations are given in 
Appendix B  (\ref{Astate} - \ref{T2zstatehalf}).
Here we follow a slightly different approach to
the same result: instead of representing $H^{{\rm Tb}}_{\rm eff}$ (\ref{HTbeff})
and $H_{\rm{ex}}$ (\ref{Hex})
in the basis given by (\ref{Astate} - \ref{T2zstatehalf}), we use the
basis $\left|\pm\pm\pm\pm\right\rangle_{\frac{1}{2}}$ for the spin-1/2 
states and the corresponding 
Tb$^{3+}$ tetrahedron states determined by (\ref{eq:map}).
The two approaches differ in the following way.  When the 
basis (\ref{Astate} - \ref{T2zstatehalf}) is used, matrix elements 
of the Hamiltonian 
are
real, while (two of) the basis functions are complex.  
When the $|\pm\pm\pm\pm\rangle$ basis is used, the basis functions
are real but the matrix elements of the Hamiltonian are complex.   When the map
(\ref{eq:map}) is applied, the Hamiltonian matrix must be complex-conjugated.

\subsection{Matrix representation of the spin-1/2 exchange Hamiltonian $H_{\rm ex}$}
In the spin-1/2 case the following operators are replaced by the matrices:
\begin{eqnarray} 
J_z & \rightarrow & \frac{1}{2} \left(\begin{array}{cc}
1 & 0 \\ 0 & -1 \end{array} \right) \equiv \frac{1}{2}\sigma_z  \label{jz} \\
J_{+} & \rightarrow & \left(\begin{array}{cc}
0 & 1 \\ 0 & 0 \end{array} \right) \equiv S_{+}  \label{jplus} \\
J_{-} & \rightarrow & \left(\begin{array}{cc}
0 & 0 \\ 1 & 0 \end{array} \right) \equiv S_{-}.  \label{jminus} 
\end{eqnarray}
The product of operators associated with different sites
is replaced by the Kronecker product of matrices.
In this way, both 
the full spin-1/2 exchange Hamiltonian $H_{\rm ex}$~(\ref{Hex}) and
the spin-1/2 independent tetrahedra Hamiltonian
$H^A$ (\ref{HA}) can be expressed as matrices by simply using the 
replacements (\ref{jz}-\ref{jminus}).

\subsection{Matrix representation of the Tb$^{3+}$ exchange Hamiltonian
$H_{\rm eff}^{\rm Tb}$}
In order to compare the matrix representation of the 
Tb$^{3+}$ exchange Hamiltonian
to the matrix representation of the spin-1/2 exchange Hamiltonian,
the basis states
$|\pm\pm\pm\pm\rangle_{\rm Tb}$
that are used to generate the Tb$^{3+}$ matrix 
must match the order and relative phase of the basis states $|\pm\pm\pm\pm\rangle_{\frac{1}{2}}$
used to generate the spin-1/2 matrix, 
{\em i.e.}, they
must be ordered and signed according to (\ref{eq:map}) and the resulting matrices 
must be complex-conjugated.
The results 
are then expressed in 
terms of the spin-1/2 matrices
$\sigma_z$ and $S_{\pm}$.
After following this procedure, we find that 
the various Tb$^{3+}$ operators which appear in
H$_{\rm eff}^{\rm Tb}$~(\ref{HTbeff}) are represented by the following spin-1/2 matrices: 
\begin{eqnarray}
PJ_z P & \rightarrow & - j_1 \sigma_z \label{PJzP} \\
PJ_{\pm}P & \rightarrow & 0 \\
PJ_z\frac{Q}{a}J_zP & \rightarrow & -\frac{j_3^2}{\Delta}  \\
PJ_{+} \frac{Q}{a} J_{+} P & \rightarrow & 0 \\
PJ_{+}\frac{Q}{a}J_{-}P & \rightarrow & -\frac{t^2}{2\Delta}(1+\sigma_z) \\
PJ_{-}\frac{Q}{a} J_{+}P & \rightarrow & -\frac{t^2}{2\Delta}(1-\sigma_z) \\
PJ_{1z}\frac{Q}{a} J_{1+} P  & \rightarrow & -\frac{j_3 t}{\Delta} S_{1-}\sigma_{2z}\sigma_{3z}\sigma_{4z} \\
& = &  PJ_{1+}\frac{Q}{a} J_{1z}P \\
PJ_{1z}\frac{Q}{a} J_{1-} P  & \rightarrow & -\frac{j_3 t}{\Delta} S_{1+}\sigma_{2z}\sigma_{3z}\sigma_{4z}\\
&  = & PJ_{1-}\frac{Q}{a} J_{1z}P.  \label{PJQJP}
\end{eqnarray}
The $2\times 2$ identity matrix is assumed when no matrix is given.

\subsubsection{Exchange interaction for Tb$^{3+}$ spins in the independent
tetrahedra model}
Using the substitutions (\ref{PJzP}-\ref{PJQJP}), 
the independent tetrahedra exchange Hamiltonian for Tb$^{3+}$ spins
$H^{A,{\rm Tb}}_{\rm eff}$ (\ref{HATbeff}) can be expressed as a matrix.  By direct comparison
to the matrix representation of the spin-1/2 independent tetrahedra 
Hamitonian $H^A$, the following map between the exchange constants
of spin-1/2 and the Tb$^{3+}$ independent tetrahedra
models can be inferred and previous results\cite{curnoe2013} are reproduced:
\begin{eqnarray}
{\cal J}_1 &=& 4 {\cal I}_1 j_1^2 + \frac{(4 {\cal I}_1 j_1 j_3)^2}{3\Delta}
-\frac{(4{\cal I}_2 j_1 t)^2}{3\Delta} + \frac{({\cal I}_3 t^2)^2}{3 \Delta}
\nonumber \\
& & - 
\frac{({\cal I}_4 t^2)^2}{12 \Delta} \label{J1} \\
{\cal J}_2 & = & -\frac{4 {\cal I}_1 {\cal I}_2 j_1^2 j_3 t}{3 \Delta} \label{J2} \\
{\cal J}_3 & = & \frac{2 ({\cal I}_2 j_3 t)^2}{3 \Delta} - \frac{{\cal I}_1 {\cal I}_3 j_3^2 t^2}{3 \Delta} \\
{\cal J}_4 & = & \frac{(2 {\cal I}_2 j_3 t)^2}{3\Delta} 
+\frac{{\cal I}_1 {\cal I}_4 j_3^2 t^2}{3\Delta} .
\label{J4}
\end{eqnarray} 
A constant offset was also found:
\begin{eqnarray}
{\cal C} & = & -\frac{(2 {\cal I}_1 j_1 j_3)^2}{3\Delta}
-\frac{({\cal I}_1 j_3^2)^2}{3\Delta} - \frac{2 (2 {\cal I}_2 j_1 t)^2}{3 \Delta} 
 \nonumber \\
& & -\frac{(2 {\cal I}_2 j_3 t)^2}{3 \Delta} 
 - \frac{({\cal I}_3 t^2)^2}{6 \Delta} 
-\frac{({\cal I}_4 t^2)^2}{24 \Delta} .
\label{C}
\end{eqnarray}

\subsubsection{The full lattice exchange interaction for Tb$^{3+}$ spins}
We now consider the full exchange model $H^{\rm Tb}_{\rm eff}$ (\ref{HTbeff}) 
for Tb$^{3+}$ spins.  
We shall show that $H^{\rm Tb}_{\rm eff}$ is equivalent to the spin-1/2
exchange Hamiltonian $H_{\rm ex}$ (\ref{Hex}) {\em plus} additional next-nearest-neighbour
and fourth order in ${\vec J}$ interactions.

Both of the operators $H^{A,{\rm Tb}}$ and $H^{B,{\rm Tb}}$ 
appear
in the expression for $H^{\rm Tb}_{\rm eff}$, which is
expanded as:
\begin{eqnarray}
H_{\rm eff}^{\rm Tb}&  = &  PH^{A,{\rm Tb}}P + PH^{A,{\rm Tb}} \frac{Q}{a} H^{A,{\rm Tb}}P
+ \nonumber \\
& & PH^{B,{\rm Tb}}P+ 
 PH^{B,{\rm Tb}} \frac{Q}{a} H^{B,{\rm Tb}}P
+\nonumber \\
& & PH^{A,{\rm Tb}} \frac{Q}{a} H^{B,{\rm Tb}}P
+PH^{B,{\rm Tb}} \frac{Q}{a} H^{A,{\rm Tb}}P. \nonumber \\
\label{HTbeff2}
\end{eqnarray}
The first two terms were already considered in the discussion of 
independent tetrahedra model $H^{A,{\rm Tb}}_{\rm eff}$ (\ref{HATbeff}), and they correspond to the term $H^A$ 
in $H_{\rm ex}$ (\ref{Hex2}).  The 
third and fourth terms correspond to $H^B$ in $H_{\rm ex}$. 
It is obvious from symmetry considerations that the constants of the
effective spin-1/2 model for $H^B$ (\ref{HB}) should be the same as those 
found for $H^A$ (\ref{HA}).  
The sum of the first four terms therefore corresponds to $H_{\rm ex}$ (\ref{Hex},\ref{Hex2}), 
with effective coupling constants as given by (\ref{J1}-\ref{C}).

\begin{figure}
\includegraphics[height=2.6in]{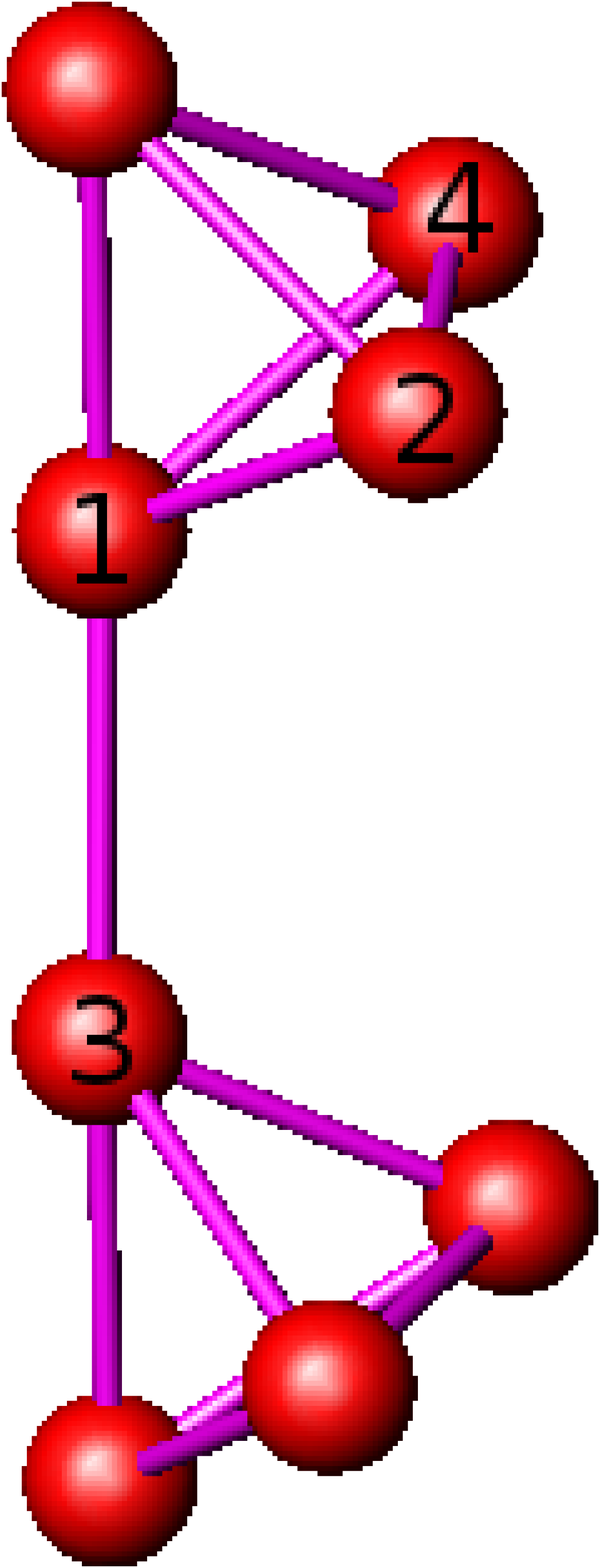}
\includegraphics[height=2.6in]{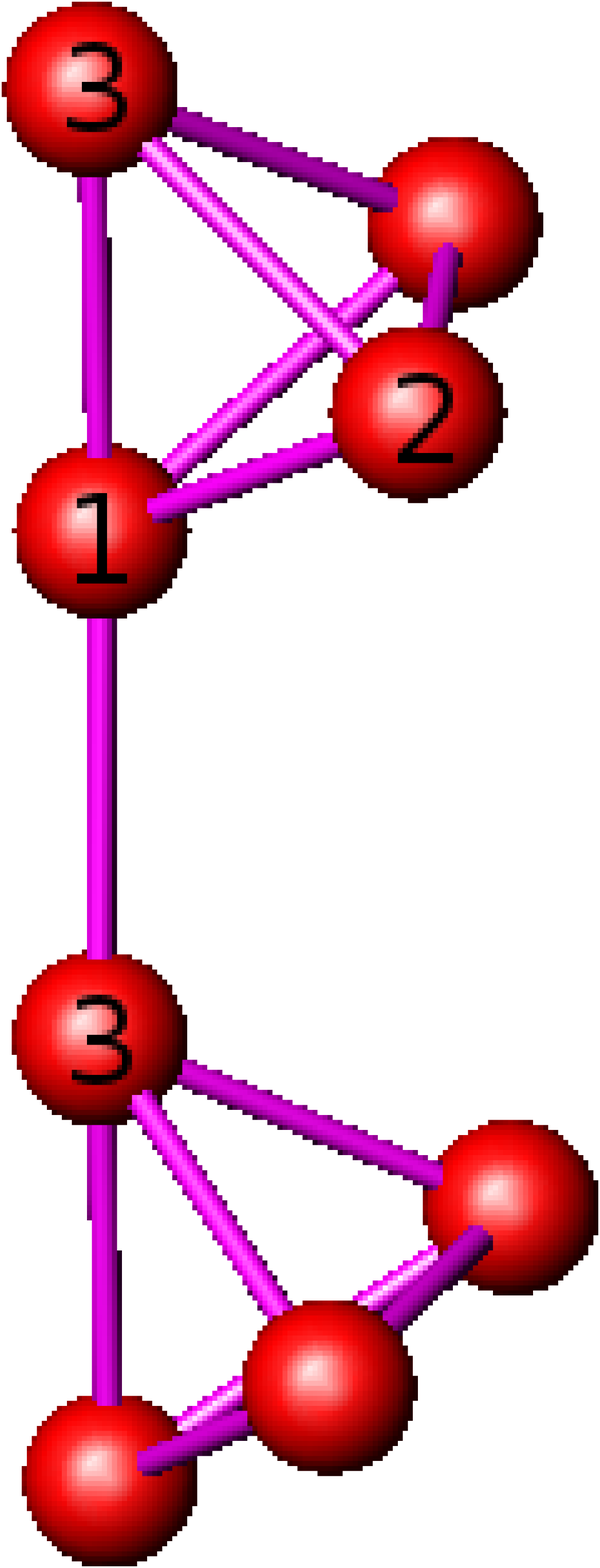}
\caption{(Colour online) Examples of arrangements of ions
involved in fourth order exchange terms. The 
left diagram corresponds to $X'_7$ and the right
to $X'_8$.}
\label{temp}
\end{figure}

The last two terms of (\ref{HTbeff2}) yield additional
next-nearest-neighbour (n.n.n.) interactions, 
and some unusual fourth order (in $\vec J$) interactions.
Symmetry considerations  determine the most general form of
these interactions.
When the two interacting spins are at different site numbers
then the interaction takes the same general form as $H_{\rm ex}$ (\ref{Hex},\ref{X1}-\ref{X4}), except
that the sum is over pairs of next-nearest neighbours with 
different site numbers,
$s_i \neq s_j$.  
These four contributions will be denoted $X_i'$, $i=1,2,3,4$.
In addition, there are two interaction terms between spins 
that are next-nearest-neighbours with the same site number ($s_i = s_{i'}$), for 
a total of 6 n.n.n. exchange coupling constants:
\begin{eqnarray}
H_{\rm ex, n.n.n.} & = & {\cal J}_1' X_1' + 
{\cal J}'_2 X_2' + {\cal J}_3' X_3' + {\cal J}_4' X_4'
+ \nonumber \\
& & {\cal J}'_5 X_5' + {\cal J}_6' X_6'
\label{Hexnnn}
\end{eqnarray}
where
\begin{eqnarray}
X_5' & = & \sum_{\langle\langle i,i'\rangle\rangle} J_{iz} J_{i'z}  \;\;
\mbox{($s_i=s_{i'}$)}\\
X_6' & = & \frac{1}{2} \sum_{\langle\langle i,i'\rangle\rangle}  J_{i+}J_{i'-} + J_{i-}J_{i'+}
\;\;
\mbox{($s_i=s_{i'}$)}.
\end{eqnarray}
Among the many different fourth order in $\vec J$ 
terms which may appear in $H_{\rm eff}^{\rm Tb}$, 
the ones that are produced by the last two terms of (\ref{HTbeff2}) are
\begin{equation}
H_{\rm ex, 4-order}  =  {\cal J}'_7 X_7' + {\cal J}'_8 X'_8 
\label{Hex4-order}
\end{equation}
where
\begin{eqnarray}
X_7' &=& \sum_{\langle i,j,k,l \rangle}  \Lambda_{ijkl} J_{i+}J_{jz}J_{kz}J_{lz}+ J_{i-}J_{jz}J_{kz}
J_{lz} \\
X_8' & = & \sum_{\langle i,j,l,i'\rangle}  \Lambda_{ijki} J_{i+}J_{jz}J_{kz}J_{i'z} .
\end{eqnarray}
In $X_7'$, the sites $i,j,k,l$ have
different site numbers.  Three of the sites ($i$, $j$, $k$) form a triangle, 
and the 
fourth site $l$ is connected to the 
triangle by a nearest neighbour bond (but it 
does not complete a tetrahedron).
In $X_8'$, the sites $i,j,k$ have different site numbers.   They are 
arranged in a triangle, while the fourth site $i'$ has 
the same site number as the site $i$ and is connected to the triangle
by a nearest neighbour bond.
Examples of arrangements of ions involved in these interactions are shown in 
Fig.\ \ref{temp}.
Similar to the expressions for $X_2$ and $X_3$, 
the phases $\Lambda_{ijkl}$  are fixed by symmetry considerations
with values of 
either 1, $\varepsilon$ or $\varepsilon^2$.

Matrix representations for $H_{\rm ex, n.n.n.}$
(\ref{Hexnnn}) and $H_{\rm ex,4-order}$
(\ref{Hex4-order}) can be written using the substitutions (\ref{jz}-\ref{jminus}).
By comparing these with the matrix representation of $H_{\rm eff}^{\rm Tb}$,
the constants ${\cal J}_i'$ can be inferred:
\begin{eqnarray}
{\cal J}_{2,3,4,6}' & = & 0\\
{\cal J}_1' & = &  \frac{2 (2 {\cal I}_1 j_1 j_3)^2}{3 \Delta}
- \frac{2 (2 {\cal I}_2 j_1 t)^2}{3 \Delta}\\
{\cal J}_5' & = & - \frac{2 (2 {\cal I}_1 j_1 j_3)^2}{9 \Delta}
-\frac{(4 {\cal I}_2 j_1 t)^2}{9 \Delta}\\
{\cal J}_7' & = & \frac{32 \sqrt{2} {\cal I}_1 {\cal I}_2 j_1^2 j_3 t}{9 \Delta}\\
{\cal J}_8' & = & -\frac{16 \sqrt{2} {\cal I}_1 {\cal I}_2 j_1^2 j_3 t}{9 \Delta
}.
\end{eqnarray}

In summary, we have considered two different exchange 
models for Tb$^{3+}$ spins in \tb, the independent tetrahedra model
and the full lattice model.  
The maps for these models onto spin-1/2 exchange models can be 
illustrated schematically as
\\
\\
\begin{tabular}{lcl}
\tb\   &   &  spin-1/2  \\
independent tetrahedron &$\rightarrow$ & independent tetrahedron \\
anisotropic exchange   & & anisotropic exchange  \\
\\
 &  & spin-1/2 \\
\tb\  & & full lattice \\
full lattice  & $\rightarrow$ & anisotropic exchange \\ 
anisotropic exchange & & plus next-nearest neighbour \\
& & and four-body interactions.
\end{tabular}
\\
\\
The relation between the anisotropic nearest-neighbour exchange constants, 
given by 
(\ref{J1}-\ref{J4}), is the same for both models.

\section{Discussion}
Recent magnetisation measurements on \tb\ have been performed by a few groups
[\onlinecite{legl2012,lhotel2012,sazonov2013}].
In the presence of a magnetic field, the Hamiltonian for \tb\ is
\begin{equation}
H({\vec B}) = H_{\rm CEF } + H_{\rm ex}^{\rm Tb} + \sum_i \mu_B g_{J} \vec{J}_i\cdot \vec{B}
\end{equation}
where $g_J = \frac{3}{2}$ is the Land\'e $g$-factor
for Tb$^{3+}$ and $\mu_B$ is the Bohr magneton.
Being unsolvable, $H({\vec B})$ is normally handled using either a 
self-consistent
mean field approximation or by using the independent tetrahedron model.
Both of these methods involve considerable simplifications of 
$H(\vec B)$.  In the former, nearest neighbour exchange interactions are replaced 
by an effective mean field, and the problem is reduced to the solution of a 
single ion Hamiltonian.  In the latter, a single tetrahedron is solved
but correlations between tetrahedra are omitted.  

%The magnetisation per atom, in units of $\mu_B$, is
%\begin{equation}
%\vec M  =  \frac{g_J}{4} \sum_{i=1}^4 \langle \vec J \rangle .
%\end{equation}
%Fits were performed using the magnetisation data in Ref.\ \onlinecite{lhotel2012} to find the four exchange constants of the independent tetrahedron model.  The matrix elements $j_1$, $j_2$, $j_3$ and $t$, given by Eqs.\ (\ref{j1}-\ref{t}) were input parameters in the fit.  The results are shown in Fig.\ \ref{mag-fit}.  The constants found {\em for the indepedent tetrahedra model} are (in kelvin)
%\begin{eqnarray}
%{\cal I}_1 & = & -1.5 \label{magnetA1} \\
%{\cal I}_2 & = &  .13 \\
%{\cal I}_3 & = & -.4\\
%{\cal I}_4 & = & -1.8 \label{magnetA4}
%\end{eqnarray}
%
%The same data was used to perform a similar fit, except that matrix 
%elements derived 
%from the CEF states in Ref.\ \onlinecite{zhang2014} were used in the 
%fit.  The matrix elements are: $j_1 = 2.61$, $j_2=-1.29$, $j_3=3.43$ and
%$t=4.83$.
%The exchange constants for the indepedent tetrahedra model 
%determined from this fit  are 
%\begin{eqnarray}
%{\cal I}_1 & = &  \label{magnetB1} \\
%{\cal I}_2 & = &  \\
%{\cal I}_3 & = & \\
%{\cal I}_4 & = & . 
%\label{magnetB4}
%\end{eqnarray}
%The plot of the fit is indistinguishable from Fig.\ \ref{mag-fit}.
%
Using the mean field approach, 
approximate values for the exchange constants for 
\tb\ were obtained.\cite{sazonov2013}  
The relation between the exchange constants used
in Ref.~\onlinecite{sazonov2013} and those defined by (\ref{Hex},\ref{X1}-\ref{X4})  is given in Appendix~C.
Using our definitions,
the constants are (in kelvin) 
\begin{eqnarray}
{\cal I}_1 & = & -.128 \label{bonville1} \\
{\cal I}_2 & = & -.083 \\
{\cal I}_3 & = & -.1595\\
{\cal I}_4 & = & -.281
\label{bonville4}
\end{eqnarray}
It should be noted that in  Ref.\ \onlinecite{sazonov2013} a constraint was applied 
in determining these numbers (it was assumed that the anti-symmetric exchange
term was absent), such that in effect only three parameters within
the four parameter space were explored.   
Nevertheless, these numbers can provide estimates of the 
exchange constants for the effective spin-1/2 model.
Ref.\ \onlinecite{sazonov2013} uses CEF states derived from the CEF Hamiltonian in
Ref.\ \onlinecite{mirebeau2007}, with matrix elements $j_1= -3.4$, $j_2=4.3$, $j_3=-2.0$, $t=4.65$ (defined by Eqs.\ (\ref{j1}-\ref{t}))
and $\Delta=18.24$ K.  Among all the exchange constants for the 
spin-1/2 model, only ${\cal J}_1$ has a first order in perturbation theory
correction; the rest are non-zero only in second order.  Using Eq.\ \ref{J1},
and the numbers given above, we calculate ${\cal J}_1 \approx -6.1$. 
The other constants are calculated using (\ref{J2}-\ref{J4}),
which yields
 ${\cal J}_2 \approx .085$, ${\cal J}_3 \approx -0.011$
and  ${\cal J}_4 
\approx 0.10$; however, without more accurate knowledge of the
\tb\ exchange constants, these values are likely not very meaningful.
The values obtained in Ref.\ \onlinecite{curnoe2013}, also highly approximate,
are in rough agreement, ${\cal J}_1 \approx -5.1$, ${\cal J}_2 \approx 0.2$, ${\cal J}_3 \approx
0.1$ and 
${\cal J}_4 \approx 0.3$.
%The most important features of the calculated ${\cal J}_1$ are its sign, which is the same as the sign on ${\cal I}_1$ (unless ${\cal I}_1$ happens to be very small), and the magnitude, which is proportional to $j_1^2$.

The negative sign of ${\cal J}_1$ indicates that \tb\ is a spin-1/2 spin ice,
with quantum fluctuations arising from the other terms in $H_{\rm ex}$,
in agreement with recent observations of spin ice-like correlations
in \tb.\cite{fennell2012,fritsch2013}
It will be interesting to see how magnetic monopoles, which are
postulated to exist as excitations in spin ices,\cite{castelnovo2008}
may be manifested in \tb.

%Our results may be compared to the constants determined for \yb: ${\cal J}_1 = -6$, ${\cal J}_2 = -3.45$, ${\cal J}_3 = 1.7$ and ${\cal J}_4 = 3.9$.  In \tb, ${\cal J}_{2,3,4}$  are non-zero only in second order perturbation theory, which accounts for their relative smallness compared to the constants of \yb.  Both materials may be classified as ``quantum spin ices" because of the sign on ${\cal J}_1$ and the presence of non-zero ${\cal J}_{2,3,4}$, however 

Using either set of estimates for the exchange constants given above
to locate the position of \tb\
in the phase
diagrams presented in Refs.\ \onlinecite{savary2012a} and \onlinecite{lee2012},
the ground state of \tb\ is predicted to be a ``quantum spin liquid" (QSL), or possibly a
``Coulomb ferromagnet" (CFM) close to the QSL boundary.
Both of these are highly entangled quantum mechanical states, with the CFM state
distinguishable from the QSL state by a non-zero magnetisation.
However, a complete description of \tb\ is almost certainly more complicated 
due to interactions with lattice structure\cite{maczka2008,bonville2014} or 
elastic strain.\cite{aleksandrov1981,aleksandrov1985,mamsurova1986,ruff2007,ruff2010,luan2011,nakanishi2011,fennell2014}

\section{Summary}
Symmetry-based analysis (group theory) is a powerful means of
reducing the complexity of 
highly symmetric crystals with limited degrees of freedom.
The observation that non-Kramers doublets and a spin-1/2 spinors
possess the same symmetry when considered in groups of four (the four 
vertices of a tetrahedron in the pyrochlore lattice) 
is unexpected, non-trivial, and very useful.
It defines a map between non-Kramers Tb$^{3+}$ and spin-1/2 basis states,
which in turn
provides the basis for a map between the exchange interaction 
specific to \tb\ and a generic spin-1/2 model.
Furthermore, the map 
easily incorporates (via perturbation theory) the effects of 
a low-lying crystal electric field excited state.
However, in order to calculate the spin-1/2 anisotropic exchange constants
with quantitative accuracy,  precise determinations of the 
anisotropic exchange constants and the CEF Hamiltonian for
\tb\ are essential.

\begin{acknowledgments}
This work was supported by NSERC.
\end{acknowledgments}

\appendix

\section{Local axes for rare earth ions in pyrochlore crystals}

For site \#1,  the local $z$-axis
is parallel to the $[111]$ direction and the local $x$- and $y$-axes are chosen to be perpendicular to $z$ and to obey the
right
hand rule.
These local axes define a set of magnetic operators
\begin{eqnarray}
J_{1x} &=& (J_1^x+J_1^y - 2J_1^z)/\sqrt 6\\
J_{1y} &=& (-J_1^x + J_1^y)/\sqrt{2}\\
J_{1z} &=& (J_{1}^x + J_1^y + J_1^z )/\sqrt{3}
\end{eqnarray}
where subscripts are used for operators using local axes and superscripts 
for global axes.

Local axes for site \#2 are defined by rotating the \#1 axes by $C_{2z}$ (this
operation also exchanges sites \#1 and \#2):
\begin{eqnarray}
J_{2x} & = & (-J_2^x - J_2^y -2 J_2^z)/\sqrt{6}\\
J_{2y} &= & (J_2^x -J_2^y)/\sqrt{2}\\
J_{2z} &=& (-J_2^x-J_2^y + J_2^z)/\sqrt 3.
\end{eqnarray}
Similarly, local axes for site \#3 are defined by rotating the \#1 axes by
$C_{2y}$:
\begin{eqnarray}
J_{3x} &=& (-J_3^x + J_3^y + 2J_3^z)/\sqrt 6\\
J_{3y} &=& (J_3^x + J_3^y)/\sqrt 2\\
J_{3z} &=& (-J_3^x + J_3^y - J_3^z)/\sqrt 3.
\end{eqnarray}
Finally, local axes for site \#4 are defined  by rotating \#1 axes by $C_{2x}$:
\begin{eqnarray}J_{4x} &=& (J_4^x - J_4^y + 2 J_4^z)/\sqrt 6\\
J_{4y} &=& (-J_4^x - J_4^y)/\sqrt 2\\
J_{4z} &= & (J_{4}^x - J_4^y - J_4^z)/\sqrt 3.
\end{eqnarray}

\section{Tetrahedron basis functions}
A suitable set of basis functions for the non-Kramers doublet
that transform according to the 
irreducible representations $A_1\oplus 3E \oplus 2 T_1 \oplus T_2$ is\cite{curnoe2007,curnoe2013}
\begin{widetext}
\begin{eqnarray}
(|A_1\rangle & = & (|++--\rangle + |+-+-\rangle + |+--+\rangle + 
|--++\rangle +  |-+-+\rangle + |-++-\rangle)/\sqrt 6 \label{Astate} \\
|E_{+}^{(1)} \rangle & = & -|----\rangle, \; \; \;
|E_{-}^{(1)} \rangle  =  -|++++\rangle \\
|E_{+}^{(2)}\rangle & = & (|-+++\rangle +|+-++\rangle + |++-+\rangle 
+|+++-\rangle)/2 \\
|E_{-}^{(2)} \rangle & = & (|+---\rangle + |-+--\rangle + |--+-\rangle 
+|---+\rangle )/2 \\
|E_{+}^{(3)})\rangle & = & (|++--\rangle + \varepsilon^2|+-+-\rangle
+\varepsilon |+--+\rangle + |--++\rangle + \varepsilon^2|-+-+\rangle
+\varepsilon|-++-\rangle)/\sqrt 6 \\
|E_{-}^{(3)}\rangle & = &  (|++--\rangle + \varepsilon|+-+-\rangle
+\varepsilon^2 |+--+\rangle + |--++\rangle + \varepsilon|-+-+\rangle
+\varepsilon^2|-++-\rangle)/\sqrt 6 \\
|T_{1z}^{(1)}\rangle & = & (|+++-\rangle +|++-+\rangle -|+-++\rangle -|-+++\rangle +
|+---\rangle +|-+--\rangle -|--+-\rangle \nonumber \\
& &  -|---+\rangle)/2 \sqrt 2 \\
|T_{1z}^{(2)}\rangle & = & (|--++\rangle -|++--\rangle)/\sqrt 2 \\
|T_{2z}\rangle & = & -(|+++-\rangle +|++-+\rangle -|+-++\rangle -|-+++\rangle -
|+---\rangle -|-+--\rangle +|--+-\rangle \nonumber \\
& & +|---+\rangle)/2 \sqrt 2 .
\label{T2zstate}
\end{eqnarray}
\end{widetext}
The states $|T_{1,2x}\rangle$ and $T_{1,2y}\rangle$  can be found by 
rotating $|T_{1,2z}\rangle$.
The corresponding spin-1/2 tetrahedron states can be found
using (\ref{eq:map}).
They are
\begin{widetext}
\begin{eqnarray}
|A_1\rangle & = & (|++--\rangle + |+-+-\rangle + |+--+\rangle + 
|--++\rangle +  |-+-+\rangle + |-++-\rangle)/\sqrt 6 \label{Astatehalf} \\
|E_{+}^{(1)}\rangle & = & |++++\rangle, \; \; \;
|E_{-}^{(1)}\rangle  =  |----\rangle \\
|E_{+}^{(2)} \rangle & = & (|+---\rangle +|-+--\rangle +|--+-\rangle
+|---+\rangle )/2\\
|E_{-}^{(2)}\rangle & = & - (|-+++\rangle +|+-++\rangle + |++-+\rangle
+|+++-\rangle)/2 \\
|E_{+}^{(3)} \rangle & = & (|++--\rangle + \varepsilon|+-+-\rangle + \varepsilon^2|+--+\rangle +|--++\rangle + \varepsilon|-+-+\rangle + \varepsilon^2|-++-\rangle)/\sqrt 6\\
 |E_{-}^{(3)}\rangle &=& (|++--\rangle + \varepsilon^2|+-+-\rangle + \varepsilon|+--+\rangle +|--++\rangle + \varepsilon^2|-+-+\rangle + \varepsilon|-++-\rangle)/\sqrt 6\\
|T_{1z}^{(1)}\rangle & = & (|+++-\rangle +|++-+\rangle -|+-++\rangle -|-+++\rangle -
|+---\rangle -|-+--\rangle +|--+-\rangle \nonumber \\
& & +|---+\rangle)/2 \sqrt 2\\
|T_{1z}^{(2)}\rangle & = & (|++--\rangle -|--++\rangle)/\sqrt 2 \\
|T_{2z}\rangle & = & (|+++-\rangle +|++-+\rangle -|+-++\rangle -|-+++\rangle +
|+---\rangle +|-+--\rangle -|--+-\rangle \nonumber \\
& & -|---+\rangle)/2 \sqrt 2.
\label{T2zstatehalf}
\end{eqnarray}
\end{widetext}

As described
elsewhere,\cite{curnoe2013} using the spin-1/2 single tetrahedron
states, $H^A_n$ can be represented as a block matrix, with the eigenvalues 
${\cal J}_1/6-2 {\cal J}_4/3$ and $-2 {\cal J}_3/3+{\cal J}_4/6$ in the $A_1$  and $T_2$ sectors, and the matrices
$$
\left( \begin{array}{ccc}  
-\frac{{\cal J}_2}{2} & 0 & \frac{\sqrt 2 {\cal J}_3}{\sqrt 3}\\
0 & -\frac{{\cal J}_4}{2} & - \frac{2{\cal J}_2}{\sqrt 3} \\
\frac{\sqrt 2 {\cal J}_3}{\sqrt 3} &  - \frac{2{\cal J}_2}{\sqrt 3} 
& \frac{{\cal J}_1}{6} + \frac{{\cal J}_4}{3} \end{array}
\right)  \mbox{ and } 
\left(\begin{array}{cc}
\frac{2 {\cal J}_3}{3} + \frac{{\cal J}_4}{6} & \frac{2 \sqrt 2 {\cal J}_2}{3} \\
 \frac{2 \sqrt 2 {\cal J}_2}{3}  & \frac{{\cal J}_1}{6} \end{array}
\right) 
$$
in the $E$ and $T_1$ sectors.  The $E$ sector is doubly degenerate
while the  $T_1$ and $T_2$ sectors are triply degenerate.
%the eigenstates of $H_n^A$ are determined by diagonalising the block matrices.
 
\section{Alternate definitions of the exchange constants}
Several different choices of definitions of the four anisotropic 
nearest neighbour exchange constants
have appeared in the literature.  
This article uses the same definitions as in [\onlinecite{curnoe2008,mcclarty2009,dalmas2012}]
with exchange constants denoted as
${\cal J}_1$, ${\cal J}_2$, ${\cal J}_3$ and ${\cal J}_4$.  

The constants used in [\onlinecite{ross2011,savary2012a,savary2012b,lee2012}], denoted $J_{zz}$, $J_{z\pm}$,
$J_{\pm\pm}$ and $J_{\pm}$, are proportional to the constants used in this work:
\begin{eqnarray}
{\cal J}_1 & = & - 3 J_{zz} \\
{\cal J}_2 & = & 3 J_{z\pm}/\sqrt 2\\
{\cal J}_3 & = & 3 J_{\pm\pm} \\
{\cal J}_4 & = & 6 J_{\pm}.
\end{eqnarray}

The constants used in the magnetisation study by Sazonov {\em et al.} (Ref.\ \onlinecite{sazonov2013}),
denoted ${\cal J}^u$, ${\cal J}^v$, ${\cal J}^w$ and ${\cal D}$,
are defined in Ref.\ \onlinecite{petit2012}.
The relation between them and the constants used in this work is
\begin{eqnarray}
{\cal J}_1 &=& -{\cal J}^u + 2 {\cal J}^w -2 \sqrt 2 {\cal D} \\
{\cal J}_2 & = & {\cal J}^u/2 + {\cal J}^w/2 + {\cal D}/(2 \sqrt 2) \\
{\cal J}_3 & = & {\cal J}^u/2 + 3 {\cal J}^v/4 - {\cal J}^w/4 - {\cal D}/\sqrt 2\\
{\cal J}_4 & = & -{\cal J}^u + 3 {\cal J}^v/2 + {\cal J}^w/2 + \sqrt 2 {\cal D}.
\end{eqnarray}
Since ${\cal J}_i$ are reserved for spin-1/2 models, 
the constants ${\cal I}_1$, ${\cal I}_2$, ${\cal I}_3$ and ${\cal I}_4$
are used instead for \tb.
Using the results obtained in [\onlinecite{petit2012,sazonov2013}],
$({\cal J}^u,{\cal J}^v,{\cal J}^w,{\cal D}) = (-.068,-.2,-.098,0)$ K, we
calculate the results given in (\ref{bonville1})-(\ref{bonville4}).

For completeness, we also include the constants used in [\onlinecite{zhitomirsky2012,zhitomirsky2014}], denoted $J_{zz}$, $J_{z\perp}$, $J_{\perp}$ and
$J_{\perp}^a$:
\begin{eqnarray}
{\cal J}_1 & = & -3 J_{zz} \\
{\cal J}_2 & = &  - \sqrt 3 J_{z\perp}/(2\sqrt 2) \\
{\cal J}_3 & = & J_{\perp} - J_{\perp}^a/4 \\
{\cal J}_4 & = & J_{\perp} + J_{\perp}^a/2.
\end{eqnarray}

 \end{document}